\documentclass[
    ,final            
  ]
  {aipproc}

\layoutstyle{6x9}

\begin{document}

\title{X-ray Spectroscopy of the Radiation-Driven Winds of Massive
  Stars: Line Profile and Line Ratio Diagnostics}


\author{David H.\ Cohen}{ address={Swarthmore College, Department of
    Physics and Astronomy, 500 College Ave., Swarthmore, PA 19081} }

\begin{abstract}

  Massive stars drive powerful, supersonic winds via the radiative
  momentum associated with the thermal UV emission from their
  photospheres.  Shock phenomena are ubiquitous in these winds,
  heating them to millions, and sometimes tens of millions, of
  degrees. The emission line spectra from the shock-heated plasma
  provide powerful diagnostics of the winds' physical conditions,
  which in turn provide constraints on models of wind shock heating.
  Here I show how x-ray line transfer is affected by photoelectric
  absorption in the partially ionized component of the wind and how it
  can be modeled to determine the astrophysically important mass-loss
  rates of these stellar winds. I also discuss how photoexcitation out
  of metastable excited levels of helium-like ions can provide
  critical information about the location of the hot plasma in
  magnetically channeled massive star winds.

\end{abstract}

\maketitle


\section{Introduction}

Massive stars are rare but due to their tremendous power output, they
contribute a significant fraction of the starlight in our galaxy and
dominate its energetics. The intense ultraviolet radiation emitted
from their surfaces drives strong, steady outflows, referred to as
radiation-driven winds.  These winds are a hallmark of massive stars,
and are the sites of atomic and plasma processes that shape their
dynamics, their physical conditions, and their emitted spectra from
which we can diagnose their properties \cite{puls2008}.

Massive stars were presumed not to be x-ray emitters by virtue of
their lack of significant subsurface convection and the absence of an
associated magnetic dynamo. Such magnetic activity generates the x-ray
emission seen in the Sun and other low-mass stars.  It was not until
the launch of the first x-ray telescopes in the 1970s that massive
stars were realized to be ubiquitous sources of strong, soft x-ray
emission \cite{harnden1979}.  Because of the presumed lack of a dynamo
and because no correlation is seen between x-ray properties and
rotation speeds in massive stars (unlike in low-mass stars, where a
positive correlation is seen) \cite{pallavicini1981}, some other
mechanism was assumed to be responsible for the x-ray emission
observed in these massive and luminous stars.

Today, at least three distinct mechanisms are thought to contribute to
the x-ray emission of massive stars.  All three are associated with
these stars' powerful radiation-driven stellar winds. And all tap the
prodigious kinetic energy of these winds, via shock heating, to
produce the observed x-rays. All massive stars with radiation-driven
winds are subject to an instability related to the nature of the
coupling of the photospheric UV radiation field to the wind ions via
scattering in resonance lines.  This instability is quite robust, and
has been shown to grow very rapidly once the wind material is several
tenths of a stellar radius above the star's surface \cite{ocr1988}.
Briefly, the instability, referred to as the Line-Driven Instability
(LDI), is due to the Doppler de-shadowing of ions that are perturbed
to move slightly faster (slower) than their neighbors. This exposes
them to more (less) photospheric radiation, which increases
(decreases) the radiation force, increasing (decreasing) their
velocities and so on. Rapidly accelerated streams then plow into
slower moving upstream material, generating strong reverse shocks.
Numerical simulations of radiation driven winds indicate that the
instability leads to shocks with strengths of several hundred km
s$^{-1}$, which heat a small fraction of the wind to temperatures of
several million K \cite{fpp1997}.

While the line-driven instability should exist in all massive stars
with strong winds, under certain circumstances one of two other
wind-shock mechanisms can occur in some massive stars and can dominate
over the instability-driven shock heating.  In very close binary star
systems in which both components are massive stars with strong winds,
the two wind flows can interact with each other directly, leading to
strong shock-heating, with shock velocities well in excess of 1000 km
s$^{-1}$, and shock temperatures approaching 100 million K.  In the
interests of using the limited space allocated to this paper, I do not
discuss this wind-wind collision mechanism any further.  The third
wind-shock mechanism also involves the head-on collision of two fast
wind flows and the associated strong shocks and high shock
temperatures.  But in this case, the two wind flows arise from a
single star and are directed toward each other -- and their head-on
collision -- by strong, large-scale magnetic fields that have been
recently discovered on a handful of massive stars.  This Magnetically
Channeled Wind Shock (MCWS) scenario seems to explain the harder x-ray
emission in some massive stars, especially very young stars that
harbor what are likely fossil fields remaining from their formation
from collapsing, magnetized interstellar clouds \cite{bm1997}.

The rest of this paper describes how atomic processes in the winds of
massive stars can be modeled to place constraints on the LDI and MCWS
x-ray production scenarios for non-magnetic and magnetic hot stars,
respectively. The next section describes the Doppler-broadened
emission lines from the hot plasma generated by the LDI and shows how
photoelectric absorption by the bulk, cool wind component in which the
hot, x-ray emitting plasma is embedded leads to a characteristic
blue-shifted and asymmetric shape for these emission lines. These
characteristic profile shapes can be used as a diagnostic of the
star's wind mass-loss rate, which is the key parameter describing the
winds of massive stars.  The third section discusses the use of the
forbidden-to-intercombination line ratios in helium-like species as a
diagnostic of the location of the x-ray emitting plasma relative to
the star's surface.  This is a non-traditional use of this particular
line ratio diagnostic, and, as shown below, it can provide very
important constraints on the MCWS mechanism, as applied to magnetic
massive stars.

\section{X-ray Emission Line Profile Diagnostics of Massive Star Wind
  Mass-Loss Rates}

Because the x-ray emitting plasma in the unstable winds of massive
stars is embedded in the bulk, cool wind that is being driven by the
radiative momentum of the photospheric UV emission, it is expected to
have the same kinematic profile as the rest of the wind.  The thermal
line emission from this plasma -- which consists primarily of
resonance lines of He-like and H-like species of abundant low-Z
elements (C, N, O, Ne, Mg, Si, S) and of L-shell lines of Fe -- is
Doppler broadened by the bulk motion of the flow.  We note that the
typical wind velocities of several 1000 km s$^{-1}$ far exceed the
sound and thermal speeds in these winds (few 10s of km s$^{-1}$ in the
bulk, cold wind and ~100 km s$^{-1}$ for low-Z ions in the
shock-heated component of the wind). Massive star winds accelerate
relatively rapidly from the stellar surface to a terminal velocity of
typically $ v_{\infty} \approx 2500$ km s$^{-1}$ in a few stellar
radii, with a velocity profile given by $v(r) = v_{\infty}(1
-R_{\ast}/r)^{\beta}$, where $R_{\ast}$ is the stellar radius and the
parameter $\beta$ typically has a value close to unity.  But because
the line emission in the hot plasma is driven by collisional
excitation, and thus scales as the square of the wind density, the
denser -- and slower moving -- base of the wind dominates the line
profile.  This leads to intrinsic emission line profiles that have
half-widths of about 1000 km s$^{-1}$.

It was only in the last ten years, since the launch of high-resolution
x-ray spectrographs aboard the {\it Chandra} and {\it XMM-Newton}
satellite telescopes, that resolved x-ray line profiles have been
measured in massive stars.  The observed x-ray line profiles have the
expected widths of roughly half the wind terminal velocities (which
are known quite accurately from measurements of ultraviolet absorption
line profiles that arise in the cold portion of these massive star
winds). These x-ray profiles, in addition to being broad, have a
characteristic blue-shifted and skewed shape that can be understood as
the effect of photoelectric absorption of the emitted x-rays in the
bulk, cool wind component.  The opacity associated with the
photoelectric absorption is relatively constant across a given line
profile, although it does vary significantly over the spectral range
of the {\it Chandra} grating spectrometers (effectively 5 to 25 \AA).
Despite the opacity's relative constancy over each emission line, its
effect on the line is far from wavelength-independent.  This is
because the absorption column density associated with the red wing of
the line, whose photons are generated in the receding, rear hemisphere
of the stellar wind, is quite large, as the photons' paths traverse
most of the wind, including the dense, inner regions of the wind.  On
the other hand, the photons on the blue wing of the emission line are
subject to much less absorption, by a much smaller wind column
density, as they traverse only part of the front hemisphere of the
wind.  Thus, the red portion of each emission line is much weaker
(more attenuated) than the blue portion, leading to the characteristic
blue-shifted and skewed profile shapes seen in the high-resolution
{\it Chandra} and {\it XMM-Newton} spectra of massive stars.

Owocki \& Cohen \cite{oc2001} have developed a simple model of such
skewed and blue-shifted x-ray line profiles, assuming a
$\beta$-velocity law, as defined above, for the kinematics of both the
cold and hot portions of the wind, which are assumed to be smoothly
mixed together with a constant filling factor of hot gas above some
onset radius, $R_{\rm o}$, below which there is assumed to be no shock
heating of the wind.  Aside from the overall normalization, the
parameters that govern the profile shapes in this model are $R_{\rm
  o}$ and $\tau_{\ast} \equiv \kappa\dot{M}/4{\pi}R_{\ast}v_{\infty}$,
which is a fiducial optical depth (and which appears as a constant in
the detailed expression for the optical depth of the wind from an
observer to any arbitrary point in the stellar wind; here, $\kappa$ is
the atomic opacity of the unshocked wind expressed as a cross-section
per unit mass, $\dot{M}$ is the wind mass-loss rate, $R_{\ast}$ is the
radius of the star, and $v_{\infty}$ is the wind terminal velocity).
By fitting the line profile model to individual emission lines in the
x-ray spectra of massive stars, we can derive values of $R_{\rm o}$
and $\tau_{\ast}$ for each line.  The former values confirm the
predictions of the LDI scenario of a shock-onset radius about half a
stellar radius above the star's surface, while the latter values
enable us to determine the mass-loss rate of a given star's wind, if
we are able to make reasonable assumptions about the other parameters,
most importantly the atomic opacity of the wind, $\kappa$.

\begin{figure}
  \includegraphics[height=.3\textheight]{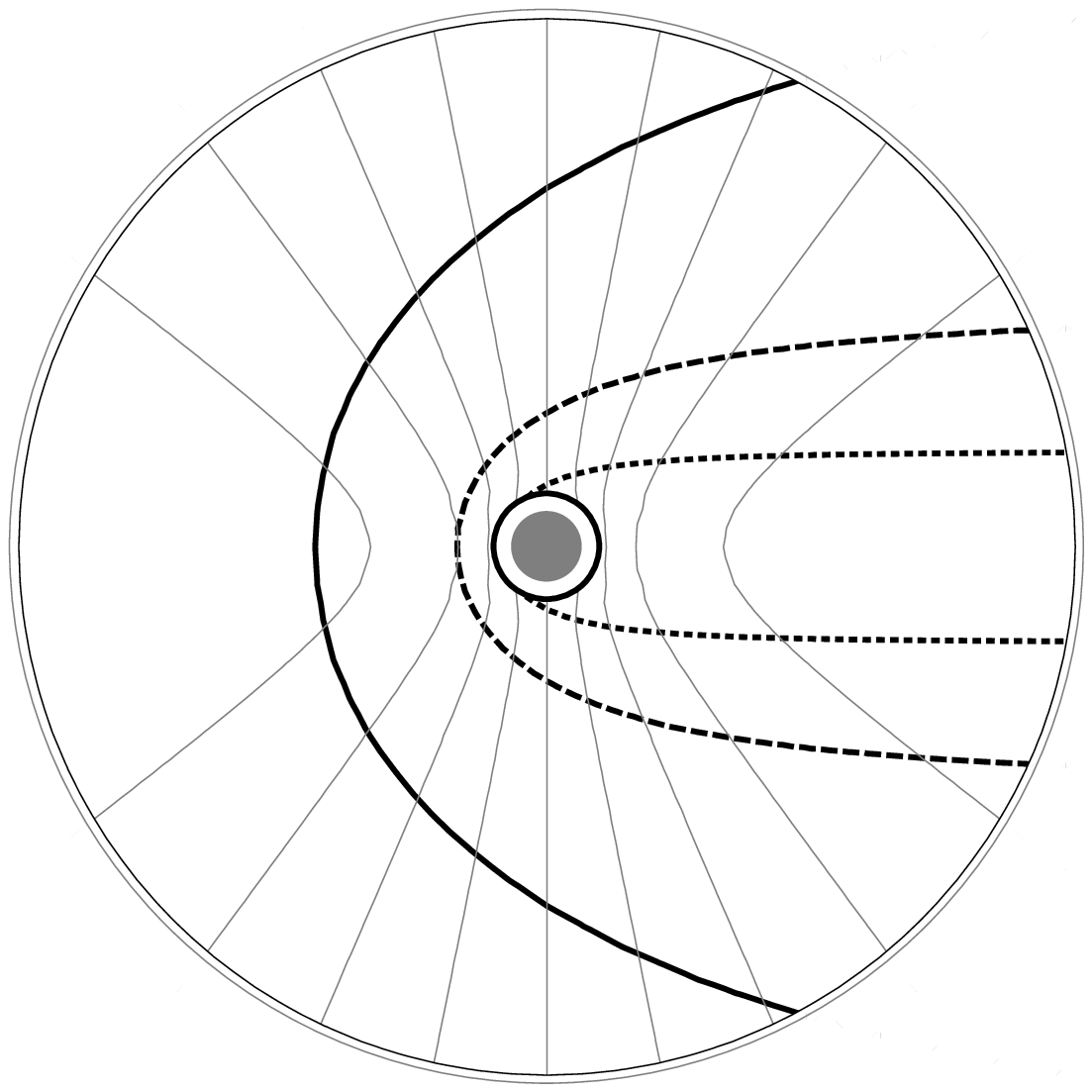}
  \includegraphics[angle=90,height=.28\textheight]{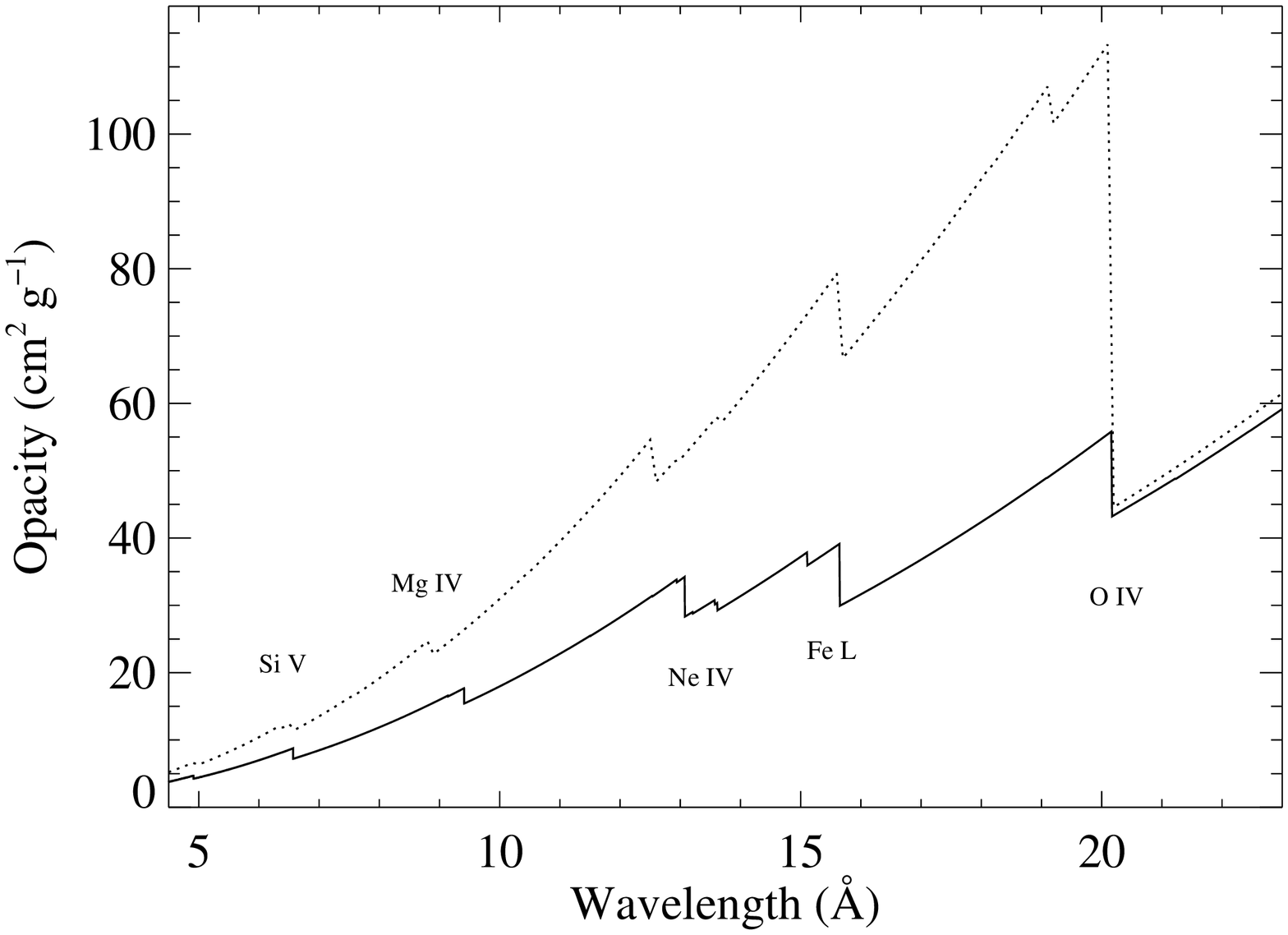}
  \caption{In the left panel is a schematic of the scenario that leads
    to the Doppler broadened and asymmetric x-ray emission line
    profiles. The observer is to the left, and the thin gray lines are
    contours of constant line-of-sight velocity (in increments of
    $0.2v_{\infty}$).  These show the Doppler shift of the emitted
    photons.  The filled gray circle at the center represents the
    star, and the white annulus around it shows the region in which
    there is assumed to be no x-ray emission (so that the circle at
    the outer edge of the annulus represents the adjustable parameter
    $R_{\rm o}$).  The three heavy lines are contours of constant
    optical depth ($\tau = 0.3, 1$ and 3 for the solid, long dashed,
    and short dashed contours, respectively). The particular model
    shown here has $R_{\rm o} = 1.5$ $R_{\ast}$ and $\tau_{\ast} = 2$.
    On the right we show two models for the wavelength-dependent
    continuum opacity of the bulk, unshocked wind.  The dotted line
    uses a solar abundance mixture of elements, while the solid line
    represents a model that assumes a particular set of elemental
    abundances measured from UV and optical spectra of the massive
    star, $\zeta$ Puppis, which has an overall subsolar heavy element
    abundance (``metallicity'') as well as nitrogen and oxygen
    abundances that are significantly altered by the nuclear
    processing that has occurred in the core of this star
    \cite{Bouret2009}. The atomic opacity is due to photoelectric
    absorption from partially ionized species, with inner shell
    absorption edges labeled.  }
  \label{fig:model}
\end{figure}

\begin{figure}
  \includegraphics[angle=90,height=.16\textheight]{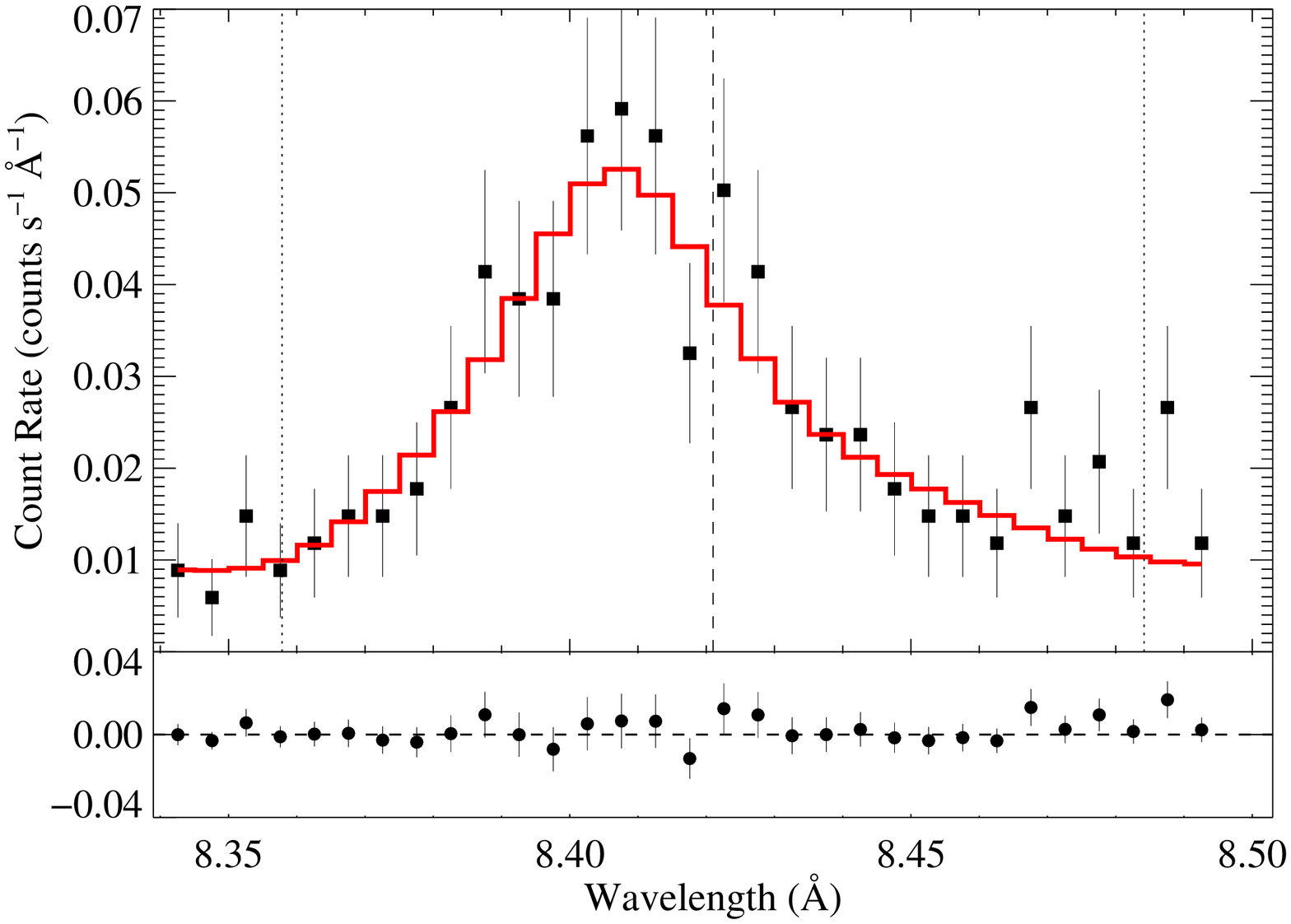}
  \includegraphics[angle=90,height=.16\textheight]{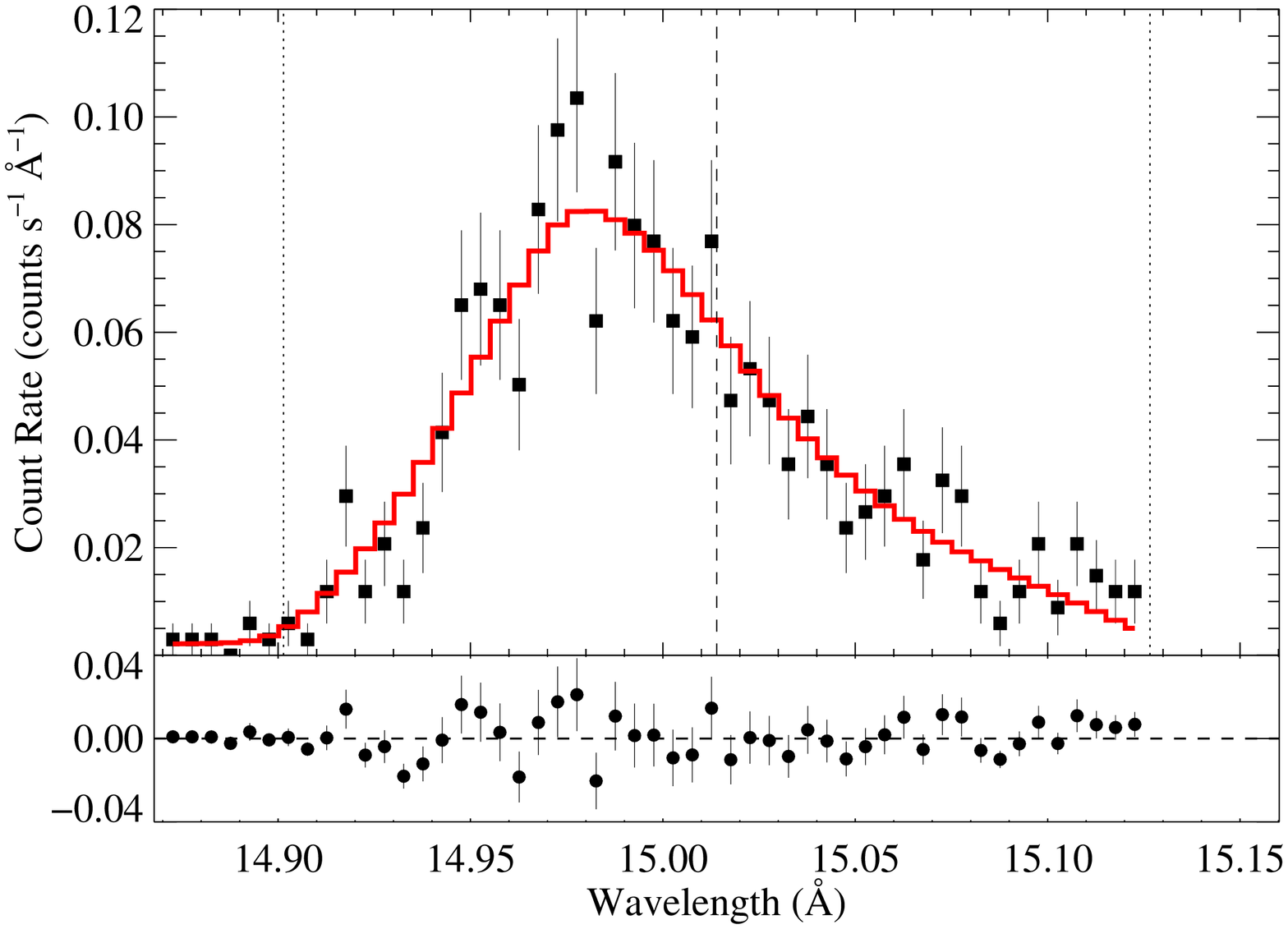}
  \includegraphics[angle=90,height=.16\textheight]{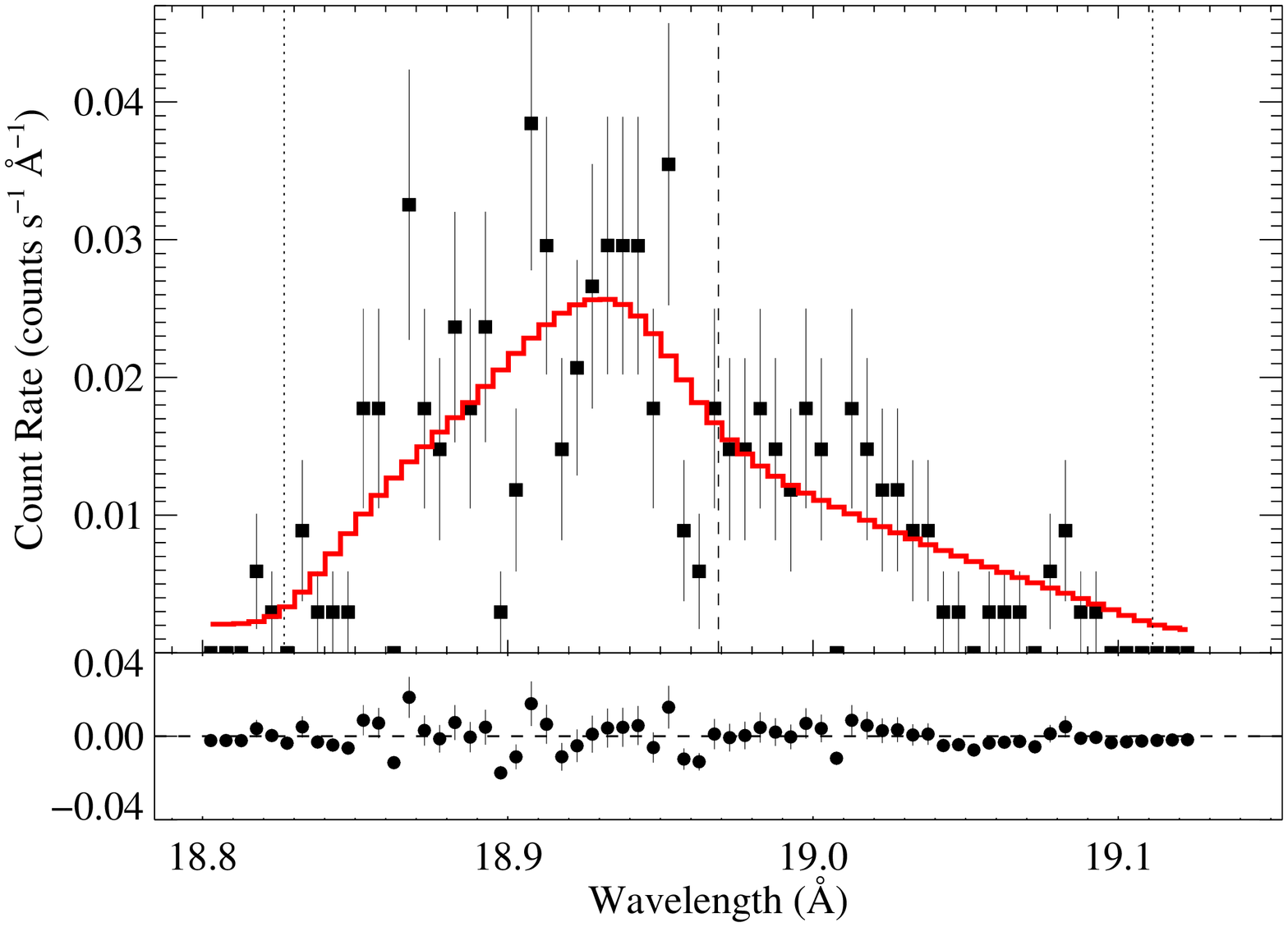}
  \caption{In these three panels are three separate emission lines
    from the {\it Chandra} spectrum of the massive star $\zeta$
    Puppis.  These are, from left to right, the Ly$\alpha$ line of
    Mg\, {\sc xii} at 8.42 \AA, the Fe\, {\sc xvii} line at 15.01
    \AA\/ which is the 3C transition from the $3d~^1{\rm P}_1$ level
    to the ground state, and the Ly$\alpha$ line of O\, {\sc viii} at
    18.97 \AA. The laboratory rest wavelengths for each emission line
    are indicated by the vertical dashed lines while the Doppler
    shifts associated with the stellar wind terminal velocity are
    represented by the light dashed vertical lines. The data are the
    points with (Poisson) error bars, the model is shown as the solid
    (red) histogram, and the fit residuals are shown in the horizontal
    windows below each data plot. The values of $R_{\rm o}$, the onset
    radius of x-ray emission we find for each line, are $R_{\rm o} =
    1.34^{+0.22}_{-0.18}$, $1.55^{+0.13}_{-0.12}$, and
    $1.18^{+0.41}_{-0.17}$ $R_{\ast}$ for the Mg, Fe, and O lines,
    respectively. This is consistent with the predictions of the LDI
    wind-shock model, in which strong shocks start to form several
    tenths of a stellar radius above the photosphere. The values we
    derive for the $\tau_{\ast}$ parameter from the profile model
    fitting are $\tau_{\ast} = 1.22^{+0.53}_{-0.45}$,
    $1.94^{+0.32}_{-0.33}$, and $3.02^{+0.52}_{-0.57}$. Note that
    these values are consistent with the wavelength dependence of the
    atomic opacity of the wind, as expected. }
  \label{fig:lines}
\end{figure}

In Fig.\ \ref{fig:model} is a schematic model of the Doppler-shifted
emission and the geometry-dependent attenuation that leads to the
characteristic broad, shifted, and asymmetric x-ray emission line
profiles seen in the high-resolution x-ray spectra of massive stars.
Also shown in this figure is a detailed model of the opacity of the
bulk, unshocked portion of the wind.  The superposition of
photoelectric absorption cross sections from multiple, abundant
elements gives the wind opacity a characteristic form that includes K-
and L-shell edges and a generally decreasing (but not monotonically
decreasing) cross section with decreasing wavelength. Because the
atomic opacity of the wind is a function of wavelength, the optical
depth to which each emission line is subject should also be a function
of wavelength. This is, indeed, the case.  In Fig.\ \ref{fig:lines}
are shown three representative emission lines from the {\it Chandra}
spectrum of the massive star $\zeta$ Puppis. I fit the line profile
model to each emission line, deriving values of $R_{\rm o}$ and
$\tau_{\ast}$ for each line. The best-fit profile models for each line
are shown along with the best-fit parameter values and their
confidence limits, listed in the caption of the figure. As can be seen
qualitatively in the figure, the longer the wavelength of the emission
line, the more shifted and asymmetric the line is.  This is borne out
from the quantitative fitting, where higher values of $\tau_{\ast}$
are seen in the longer wavelength lines.

The results shown for the three emission lines in Fig.\
\ref{fig:lines} are representative of what is found from fits to 16
lines in the {\it Chandra} grating spectrum of $\zeta$ Puppis -- the
trend of derived $\tau_{\ast}$ values with wavelength is consistent
with the wavelength dependence of the opacity.  So, with estimates of
the wind opacity, $\kappa(\lambda)$, as well as values of the star's
radius and the wind terminal velocity, an estimate of the wind
mass-loss rate of $\zeta$ Puppis can be made by fitting the ensemble
of 16 $\tau_{\ast}$ values.

\begin{figure}
  \includegraphics[angle=90,height=.3\textheight]{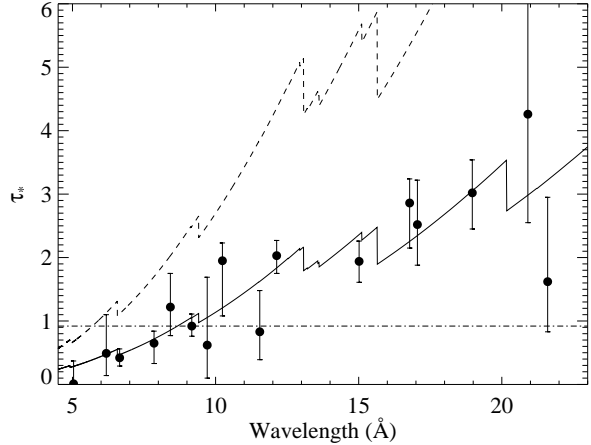}
  \caption{The points represent the best-fit $\tau_{\ast}$ values
    (along with their 68 percent confidence intervals, displayed as
    error bars) from the fitting of the line profile model to all the
    available lines in the {\it Chandra} spectrum of the massive star
    $\zeta$ Puppis.  Three of these fits were shown in Fig.\
    \ref{fig:lines}.  The horizontal dash-dotted line is the best-fit
    constant $\tau_{\ast}$ value, which does not provide a good fit to
    the data.  There is a highly statistically significant correlation
    between $\tau_{\ast}$ and wavelength.  The solid curve is the
    best-fit model based on the governing expression for $\tau_{\ast}
    \propto \kappa \dot{M}$, where the wavelength dependence arises
    from the wavelength dependence inherent to the atomic opacity,
    $\kappa$, and the mass-loss rate, $\dot{M}$, is the sole
    adjustable parameter of the fit. The upper, dashed curve
    represents the model of $\tau_{\ast}$ based on the traditional
    mass-loss rate estimate of $\dot{M} = 8.7 \times 10^{-6}$
    M$_{\odot}$ yr$^{-1}$ \cite{Repolust2004}, which is almost three
    times larger than our new estimate of $\dot{M} = 3.5 \times
    10^{-6}$ M$_{\odot}$ yr$^{-1}$. }
  \label{fig:mdot_fit}
\end{figure}

The results of this mass-loss rate fitting are shown in Fig.\
\ref{fig:mdot_fit}, where the 16 $\tau_{\ast}$ values derived from the
emission lines in the {\it Chandra} spectrum of $\zeta$ Puppis are
shown along with the model of $\tau_{\ast}(\lambda)$ that best fits
these derived $\tau_{\ast}$ values.  The best-fit value of the
mass-loss rate is $\dot{M} = (3.5 \pm 0.3) \times 10^{-6}$ M$_{\odot}$
yr$^{-1}$. This estimate of the mass-loss rate of $\zeta$ Puppis is
nearly three times lower than the traditional estimate of $\dot{M} =
8.7 \times 10^{-6}$ M$_{\odot}$ yr$^{-1}$, which was derived from
modeling the optical hydrogen Balmer-$\alpha$ emission line strength
of the star \cite{Repolust2004}. The traditional method of using
hydrogen Balmer-$\alpha$ emission as a mass-loss rate diagnostic is
problematic, however, because the strength of that emission line
scales as the square of the density of the wind (as the upper level is
populated by two-body recombination).  Thus, if density
inhomogeneities are present (i.e. ``clumping''), but not modeled, the
mass-loss rate will be overestimated.  The new method described in
this paper, using x-ray line profiles and photoelectric absorption,
which scales as the column density rather than the density-squared, is
not susceptible to this problem. Quite recently, the Balmer-$\alpha$
mass-loss rate has been re-evaluated, this time taking clumping into
account, and a value is found that is much more consistent with the
value we report here (an upper limit of $\dot{M} = 4.2 \times 10^{-6}$
M$_{\odot}$ yr$^{-1}$; as the modeling can only place a lower-limit on
the amount of clumping, and thus puts an upper limit on the mass-loss
rate) \cite{Puls2006}. So, this new technique based on modeling the
effect of photoelectric absorption in the winds of massive stars on
the x-ray line profiles from shock-heated material embedded in these
winds will provide a valuable complementary approach to determining
the mass-loss rates of massive star winds.

\section{X-ray Emission Line Ratio Diagnostics of the Location of Hot
  Plasma in Magnetically Channeled Wind Shocks}

Large scale magnetic fields are being found in massive stars with
increasing frequency as techniques for making sensitive Zeeman
splitting measurements advance \cite{Wade2008}. The canonical magnetic
massive star is $\theta^1$ Orionis C; the brightest star in the Orion
Nebula and the primary source of far UV radiation that ionizes and
illuminates the nebula. Its magnetic field is consistent with a tilted
dipole with a field strength somewhat in excess of 1 kG
\cite{Donati2002}.  An integrated picture of this star and its
magnetosphere has emerged, based on the steady-state Magnetically
Channeled Wind Shock (MCWS) model of Babel \& Montmerle \cite{bm1997}
that explains the strength, hardness, and modulation of the x-ray
emission.

The broadband x-ray properties are explained in the context of the
MCWS scenario by the magnetic channeling of the wind, which causes
wind streams from the two hemispheres of the star to collide in the
magnetic equator at significantly higher velocities than the embedded
wind shocks in the LDI model discussed in the previous section.  The
confinement of the shock heated wind plasma by the strong magnetic
field of the star explains the high levels of x-ray emission in the
MCWS scenario, as the post-shock plasma in a magnetosphere with
approximately toroidal geometry maintains a relatively high density
despite its large thermal pressure.  The rotation of the star causes
part of the magnetosphere to be periodically occulted by the star
itself. The magnetohydrodynamics simulations presented in
\cite{Gagne2005} explain the hardness and strength of the x-rays and
make a prediction about the location of the x-ray emitting plasma,
which explains the timing of the observed x-ray dimming from the
occultation of the magnetosphere induced by the star's rotation.
However, although the timing is well reproduced, the magnitude of the
observed x-ray dimming is somewhat more substantial than the MHD
simulations predict \cite{Gagne2005}.

One possible solution to this discrepancy involves having the x-ray
emitting plasma closer to the star's surface than the MHD simulations
predict. To test this, we have used the helium-like
forbidden-to-intercombination line ratio [$\mathcal{R} \equiv
z/(x+y)$, or more colloquially, $f/i$] to diagnose the distance of the
shock-heated plasma from the photosphere of $\theta^1$ Orionis C. This
line ratio, as manifest in abundant low-Z elements, traditionally has
been used as a density diagnostic, as the collisional excitation out
of the metastable $^3{\rm S}$ upper level of the forbidden line can
dominate spontaneous emission if the density is high enough
\cite{gj1969}. However, in environments with strong UV radiation,
photoexcitation out of the upper level of the forbidden line can be
more important than collisional excitation. And if one knows or can
model the emergent UV flux from the photosphere of a massive star,
then the $\mathcal{R}$ ratio can be used as a diagnostic of the
distance the X-ray emitting plasma is from the photosphere.

\begin{figure}
\vspace*{0.7in}
\hspace*{-0.25in}
 \includegraphics[scale=0.5,angle=0]{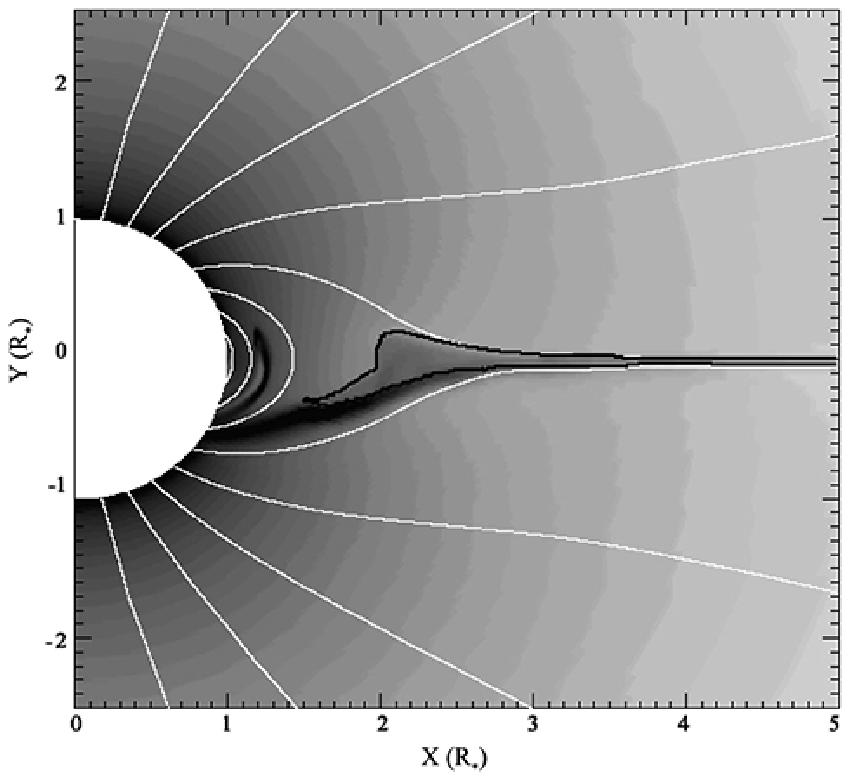} 
\vspace*{2.in}
\hspace*{-1.8in}
 \includegraphics[scale=0.5,angle=0]{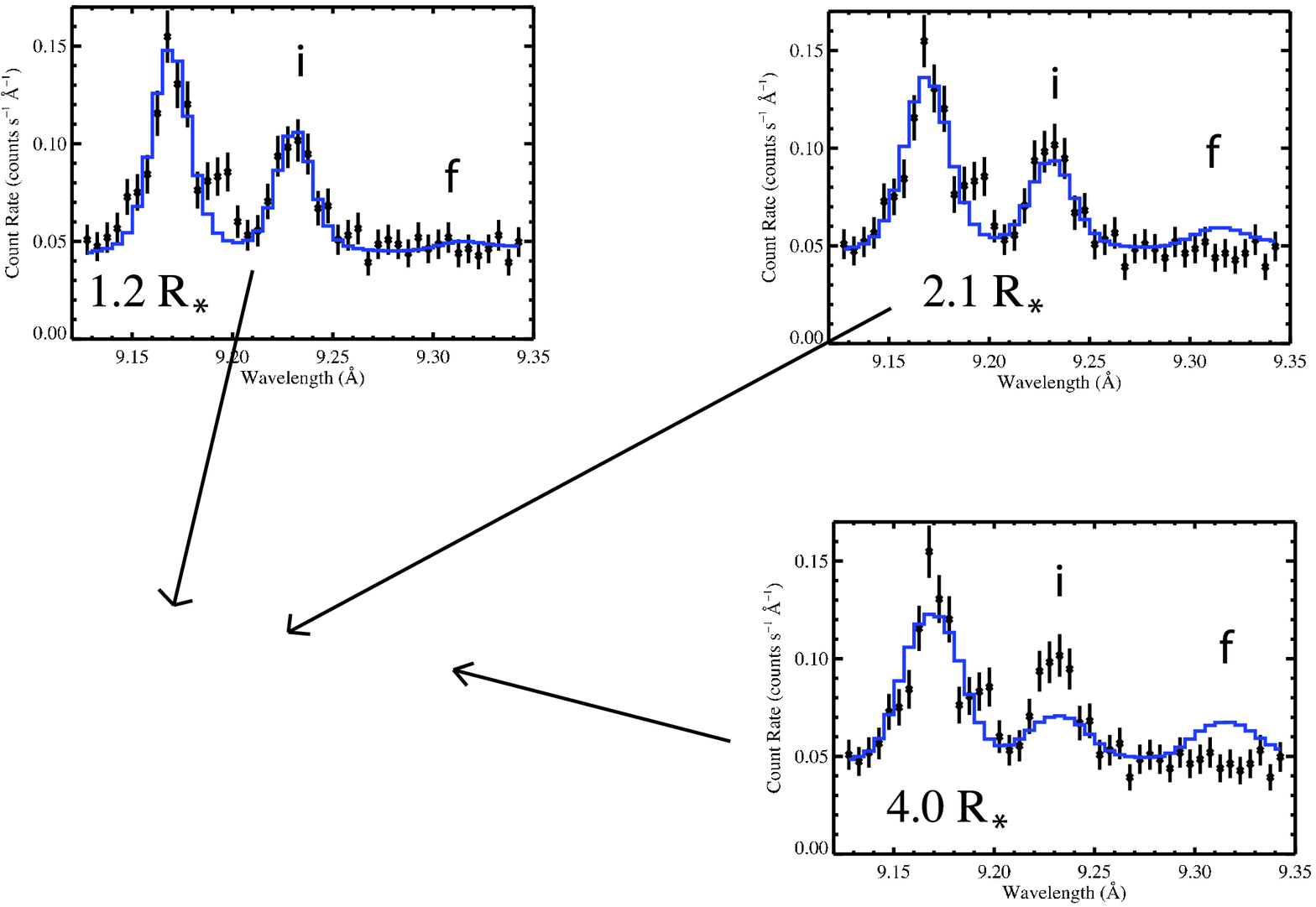} 
 \caption{In the lower left is a snapshot from a 2-D MHD simulation of
   the magnetically channeled wind of the massive star $\theta^1$
   Orionis C.  These simulations are described in \cite{Gagne2005} and
   the snapshot shown here displays the emission measure in grayscale,
   magnetic field lines as white contours, and has a thick, black
   contour enclosing plasma with temperature above $10^6$ K.  The
   three other panels show the {\it Chandra} spectrum in the vicinity
   of the helium-like Mg\, {\sc xi} complex, with a model of the
   resonance, intercombination ($i$), and forbidden ($f$) lines
   overplotted (blue histogram).  The relative strengths of the $f$
   and $i$ lines are different in each of the three panels, as the
   three models were calculated assuming a source location of 1.2
   ${\mathrm {R_{\ast}}}$, 2.1 ${\mathrm {R_{\ast}}}$, and 4.0
   ${\mathrm {R_{\ast}}}$, respectively, starting at the top left and
   moving clockwise. The arrows indicate the approximate location in
   each case. The intermediate case -- $r = 2.1~{\mathrm {R_{\ast}}}$,
   which seems to agree with the MHD simulation -- is only marginally
   consistent with the data (68\% confidence limit). }
  \label{fig:mhd}
\end{figure}

The photoexcitation mode of the $f/i$ diagnostic can be applied to the
{\it Chandra} spectrum of $\theta^1$ Orionis C.  The formalism of the
line ratio modeling is from \cite{bdt1972}, where the photoexcitation
parameter, $\phi_{\ast}$, is determined from the photospheric surface
flux given by a TLUSTY model atmosphere \cite{lh2006} and the
geometrical dilution factor, assuming optically thin radiation
transport. The best constraints are provided by the Mg\, {\sc xi}
emission line complex near 9.2 \AA, for which the forbidden line is
nearly completely suppressed by the photoexcitation effect. As can be
seen in Fig.\ \ref{fig:mhd}, which shows a representative MHD
simulation snapshot, the hot, x-ray emitting plasma in these
simulations is located near $r = 2$ ${\rm R}_{\ast}$. However, the
line ratio modeling indicates that at that distance from the
photosphere, there should be a measurable forbidden line flux in the
Mg\, {\sc xi} complex. The forbidden line does not appear to be
present in the {\it Chandra} data, however.  Formally, an upper limit
to the forbidden line flux is found that is marginally consistent with
the location of the hot plasma in the simulations.  However, smaller
radii are preferred.  These smaller radii are also more consistent
with the dimming of the x-rays at rotation phases where the
magnetosphere is expected to be occulted by the star. Perhaps future
refinements to the MHD modeling -- including extension from 2-D to 3-D
-- will lead to the bulk of the hot plasma being closer to the star's
surface in the simulations.

\section{Conclusions}

Stellar x-ray emission is generally from optically thin thermal
equilibrium coronal plasmas and, at high spectral resolution, consists
of strong lines from high ionization stages of low- and mid-Z elements
superimposed on a weak bremsstrahlung continuum.  Traditional x-ray
spectral diagnostics applied to stars include temperature
determinations from line ratios and from global spectral synthesis,
elemental abundance determinations from line ratios and
line-to-continuum ratios, and overall x-ray flux levels from the
strength of the observed x-ray emission.  In this paper I have
described two additional spectral diagnostics based on different
atomic processes and shown how they can each be applied to
high-resolution x-ray spectra of massive stars, where the x-ray
emission is produced in shock-heated stellar winds.

Photoelectric absorption causes a distinctive asymmetric and blue
shifted line shape, which arises from a purely geometric effect in a
radially expanding optically thick wind.  By quantitatively modeling
individual emission lines in the x-ray spectra of single, unmagnetized
massive stars, the continuum optical depth of the wind, due to
photoelectric absorption, can be determined.  By combining multiple
measurements at different wavelengths, a single model of the wind
optical depth, incorporating the effect of the wavelength dependence
of the continuum opacity, can be fit to the ensemble of line optical
depth values and a single wind mass-loss rate can be determined.  I
reported here on the first application of this technique to a massive
star and find, from this analysis, a lower mass-loss rate for $\zeta$
Puppis than is found from traditional mass-loss rate diagnostics.
This result is consistent with other recent and independent
reassessments of the mass-loss rate of this particular massive star.
The x-ray profile fitting mass-loss rate diagnostic is attractive
compared to other methods because it is sensitive to the column
density of the wind rather than the square of the wind density, and so
it is not subject to uncertainties due to wind clumping.

Helium-like forbidden-to-intercombination line ratios have also been
shown in this paper to be useful diagnostics of massive star winds and
the x-ray production that takes place within them.  This line ratio is
affected by photoexcitation and thus can be used as a diagnostic of
the distance that the x-ray emitting plasma is from the surface of the
star.  I have briefly shown one application of this diagnostic to a
magnetic massive star, $\theta^1$ Orionis C.  The line ratio
diagnostic applied to this particular star indicates that the MHD
models of its magnetically channeled wind may predict a location for
the x-ray producing plasma that is too far from the star's surface.

In summary, with the advent of high-resolution stellar x-ray
spectroscopy, our understanding of atomic processes as applied to the
plasma found in massive star winds enables us to diagnose important
properties of these objects and to test models of x-ray production in
them.

\begin{theacknowledgments}
  Support for this work was provided by the National Aeronautics and
  Space Administration through {\it Chandra} award number AR7-8002X to
  Swarthmore College issued by the {\it Chandra} X-ray Observatory
  Center, which is operated by the Smithsonian Astrophysical
  Observatory for and on behalf of the National Aeronautics and Space
  Administration under contract NAS8-03060. Support was also provided
  by NASA's Long Term Space Astrophysics grant NNG05GC36G and by the
  Provost's Office of Swarthmore College.
\end{theacknowledgments}



\bibliographystyle{aipproc}   

\bibliography{cohen}

\IfFileExists{\jobname.bbl}{}
 {\typeout{}
  \typeout{******************************************}
  \typeout{** Please run "bibtex \jobname" to optain}
  \typeout{** the bibliography and then re-run LaTeX}
  \typeout{** twice to fix the references!}
  \typeout{******************************************}
  \typeout{}
 }

\end{document}